\documentclass[aps,prl,twocolumn]{revtex4}
\usepackage{graphicx}
\usepackage{dcolumn}
\usepackage{tikz}
\usepackage{bm}
\usepackage{amsmath}
\usepackage{amsfonts}
\usepackage{psfrag}
\usepackage{color}

\usepackage{verbatim}

\DeclareGraphicsExtensions{.pdf,.eps}
\newcommand{\Eq}[1]{Eq.~(\ref{#1})}

\begin{document}

\title{Collective optomechanical effects in cavity quantum electrodynamics} 

\author{Erika Cortese}
\author{Pavlos Lagoudakis}
\author{Simone \surname{De Liberato}}

\affiliation{School of Physics and Astronomy, University of Southampton, Southampton, SO17 1BJ, United Kingdom}

\begin{abstract}
We investigate a cavity quantum electrodynamic effect, where the alignment of two-dimensional freely rotating optical dipoles is driven by their collective coupling to the cavity field. By exploiting the formal equivalence of a set of rotating dipoles with a polymer we calculate the partition function of the coupled light-matter system and demonstrate it exhibits a second order phase transition between a bunched state of isotropic orientations and a stretched one with all the dipoles aligned. Such a transition manifests itself as an intensity-dependent shift of the polariton mode resonance. Our work, lying at the crossroad between cavity quantum electrodynamics and quantum optomechanics, is a step forward in the on-going quest to understand how strong coupling can be exploited to influence matter internal degrees of freedom. 
\end{abstract}

\maketitle

\psfrag{OM}{$\frac{\langle\omega_-\rangle -\omega_x}{\Omega}$}
\psfrag{A}{a}

When the energy exchange between an optically active dipolar transition and a resonant electromagnetic cavity mode becomes faster than any relaxation process, we enter the so-called strong coupling regime. In such a regime a description of the system in terms of light and matter exchanging energy through emission and absorption processes fails, and  it becomes necessary to consider its coupled eigenmodes. When many independent dipoles are coupled with the same cavity photonic mode, the collective light-matter coupling scales with the square root of the number of dipoles \cite{Dicke54}, making it possible to modify the coupling by engineering the dipole density. The normal modes of those systems are called polaritons, quasi-bosonic \cite{DeLiberato09} half-light and half-matter quasi-particles. They have been observed to date in a number of different cavity quantum electrodynamics implementations, both in atomic physics, using cold atoms \cite{Tuchman06,Arnold11,Chen11,Ningyuan16,Culver16}, and in many solid-state system, from microcavity-embedded semiconductor quantum wells to magnetic spheres in microwave cavities \cite{Weisbuch92,Lidzey98,Dini03,Scalari12,Zhang14,Liu15,Gubbin16,Coles14}.

\begin{figure}[t!]
\begin{center}
\includegraphics[width=0.45\textwidth]{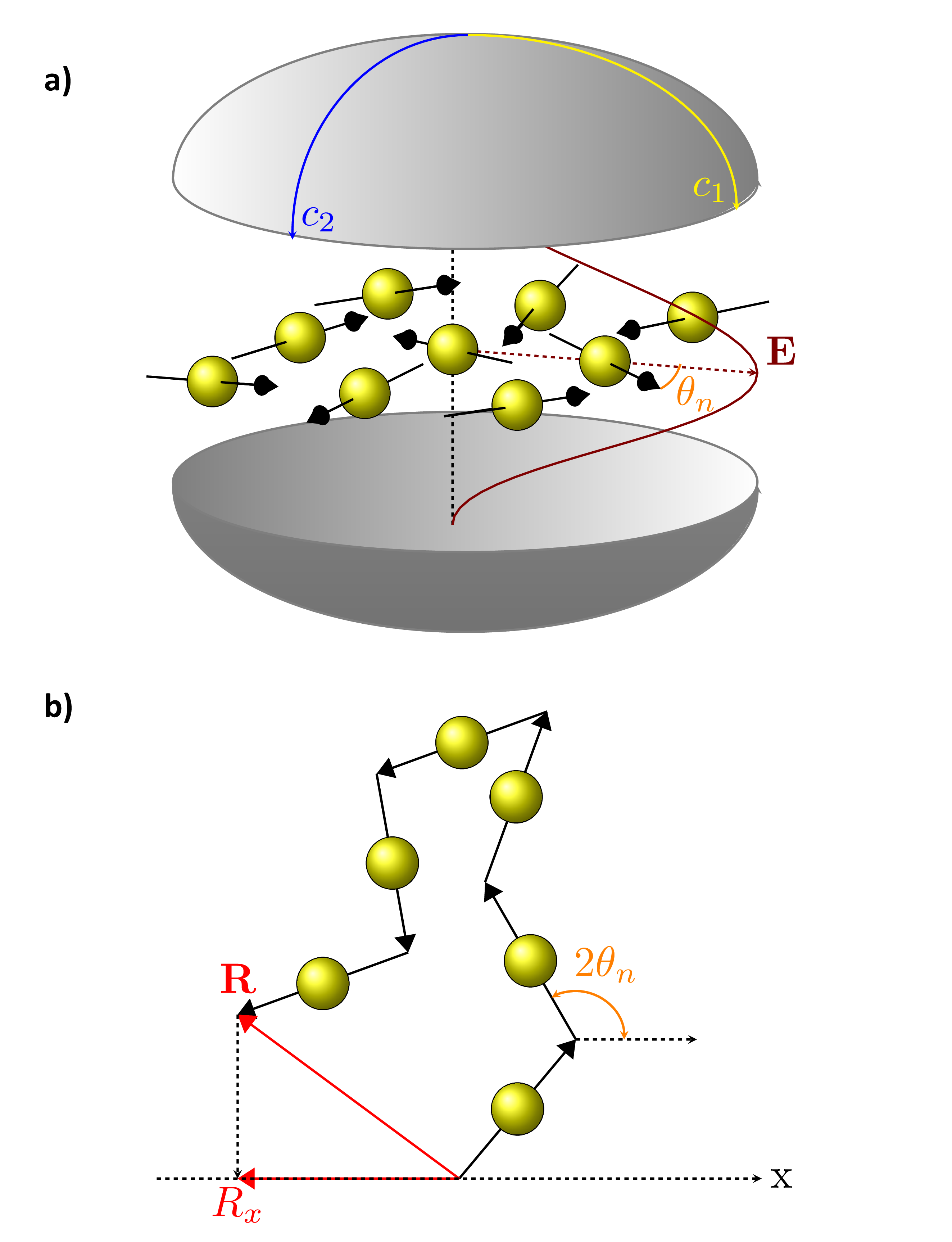}
	\caption{a) Sketch of the system under consideration: a set of rotating coplanar dipoles identically coupled to a single photonic cavity mode. The $n$th dipole forms an angle $\theta_n$ with the direction of the electric field. The cavity represented has elliptical mirrors with different curvatures along principal axis $c_1\neq c_2$, allowing to consider a single resonant mode with well defined polarization. 
	b) The phase space of the dipoles is equivalent to the one of a two-dimensional polymer, allowing us to calculate the relevant partition function.}
\label{Fig1}
\end{center}
\end{figure}

The coupling of each dipole with the photonic field will in general by influenced by the state of its microscopic degrees of freedom, like its orientation or vibrational state. Given that the energy of the light-matter coupled system depends on the total, collective coupling, the coupling to the cavity will thus generate an optomechanical force exerted on each dipole.
In most cases the light-matter coupling strength of a single dipole $\chi$ is much smaller than the temperature of the system, but the collective, enhanced coupling $\Omega\approx\sqrt{N}\chi$, with $N$ the number of coherently coupled dipoles, can easily exceed it. 
Whether the strength of the optomechanical force exerted on each dipole scales with  $\chi$ or $\Omega$ will thus determine if thermodynamical effects due to such a force are observable.
This problem has generated a remarkable interest \cite{Hutchison12,Cwik14,Shalabney15,Galego15,Cwik16,Galego16} and it has been recently addressed in two distinct theoretical works, investigating impact of strong light-matter coupling in molecular systems on underlying degrees of freedom. In particular Ref. \cite{Galego15} studies the change in bond length while Ref. \cite{Cwik16} investigates the modification due to either vibrational or rotational degrees of freedom of the coupled molecules. Both works arrive at the same conclusion, namely that while observables depending on the energy of the polariton states feel the collective coupling $\Omega$, the relevant energy scale for effects influencing internal degrees of freedom of individual molecules is the single molecule coupling $\chi$. Those effects are thus washed out by thermal fluctuations. 

In this Letter we demonstrate that this situation can be overcome in highly excited systems, in which a macroscopic density $s=\frac{M}{N}$ of excitations is present.
We will demonstrate how in this case the force felt by each dipole scales as $s\Omega$, and optomechanical effects due to the intensity of light-matter coupling can become observable in cavity quantum electrodynamics setups.

The basic intuition behind our result is that in the resonant case the energy of polariton modes is linear in $\Omega$ and their shift when a single molecule changes its internal state is thus of order of $\frac{d\Omega}{dN}\propto\frac{\chi}{\sqrt{N}}$. This shift has to be multiplied by the number of polaritons $M$, leading to a total contribution of the order of $s\Omega$. Not only this allows to reach a regime in which novel optomechanical
effects are observable, but also to explore the full dynamical range of the system using the excitation density $s$ as an effective temperature knob. 

Our aim here is to give a proof of concept for the means to harness optomechanical effects due to strong light-matter coupling for manipulating the thermodynamic properties of microscopic degrees of freedom associated with each dipole. Thus we choose to consider the simplest model presenting interesting novel physics, a set of rotating two-dimensional dipoles identically coupled to a single resonant cavity mode. The coupling to a single mode can be achieved by properly engineering the system, for example using elliptic mirrors leading to significant polarization splitting \cite{Uphoff15}, as shown in the sketch of the system presented in Fig. \ref{Fig1}a.

Similar to what done in Ref. \cite{Cwik16}, we will consider the rotational degree of freedom to be completely classical and adiabatic, that is we will assume the orientations of the different dipoles to be classical coordinates and neglect their kinetic energy.
The Hamiltonian of $N$ coplanar dipoles of frequency $\omega_x$, each one forming an angle $\theta_n$ with the electric field of the resonant cavity mode takes the form of an inhomogeneous Dicke model \cite{Cummings83,Lopez07}. It can be written, 
introducing a set of Pauli matrices $\boldsymbol{\sigma}_n$ for the $n$th molecule and the bosonic annihilation operator $a$ for the cavity photon as
\begin{eqnarray}
\label{H}
H=\omega_x a^{\dagger}a+\sum_n \left[\frac{\omega_x}{2} \sigma_n^z+
\chi \cos(\theta_n)(a\sigma^+_n
+a^{\dagger}\sigma^-_n)\right],
\end{eqnarray}
 where we work in units such that $\hbar=1$. In \Eq{H} we used the rotating wave approximation (RWA) since ultrastrong coupling effects are not expected to play any role apart from shifting the eigenfrequencies by an amount which, in the case of the polariton modes, is of order $\frac{N\chi^2}{\omega_x^2}$ and thus usually negligible.
In the bosonic regime this Hamiltonian can be diagonalised recovering $N-1$ dark modes at the bare energy $\omega_x$ \cite{Emary13,Shammah16}, and a pair of lower and upper polariton modes of energy 
\begin{eqnarray}
\label{freq}
\omega_{\pm}&=&\omega_x \pm\chi\sqrt{\sum_{n=1}^N\cos^2(\theta_n)}.
\end{eqnarray}
As expected the energies of the polariton modes now depend upon the angles $\theta_n$. 
For molecules fully oriented along the electric field ($\theta_n=0\;\forall \, n$) we have
\begin{eqnarray}
\label{wmin}
\omega_{-}^{\text{min}}&=&\omega_x-\chi\sqrt{N}=\omega_x-\Omega,
\end{eqnarray}
while if the dipoles are isotropically distributed over the plane we get the higher energy
\begin{eqnarray}
\label{wiso}
\omega_{-}^{\text{iso}}&=&\omega_x-\chi\sqrt{\frac{N}{2}}=\omega_x-\frac{\Omega}{\sqrt{2}}.
\end{eqnarray}
If the system is excited in a state containing only lower polaritons, the total energy will be minimised by having all dipoles aligned, but this effect tends to be counterbalanced by the higher entropy of non-aligned states. Our previous argument on the scaling of the relevant energy  with $s$ thus leads us to expect a phase transition, with the lower polariton energy transitioning between $\omega_{-}^{\text{min}}$ and $\omega_{-}^{\text{iso}}$ as the temperature is increased or the excitation density decreased, with a critical temperature of the order of $s\Omega$. 

\begin{figure}
\centering
	\includegraphics[width=0.5\textwidth]{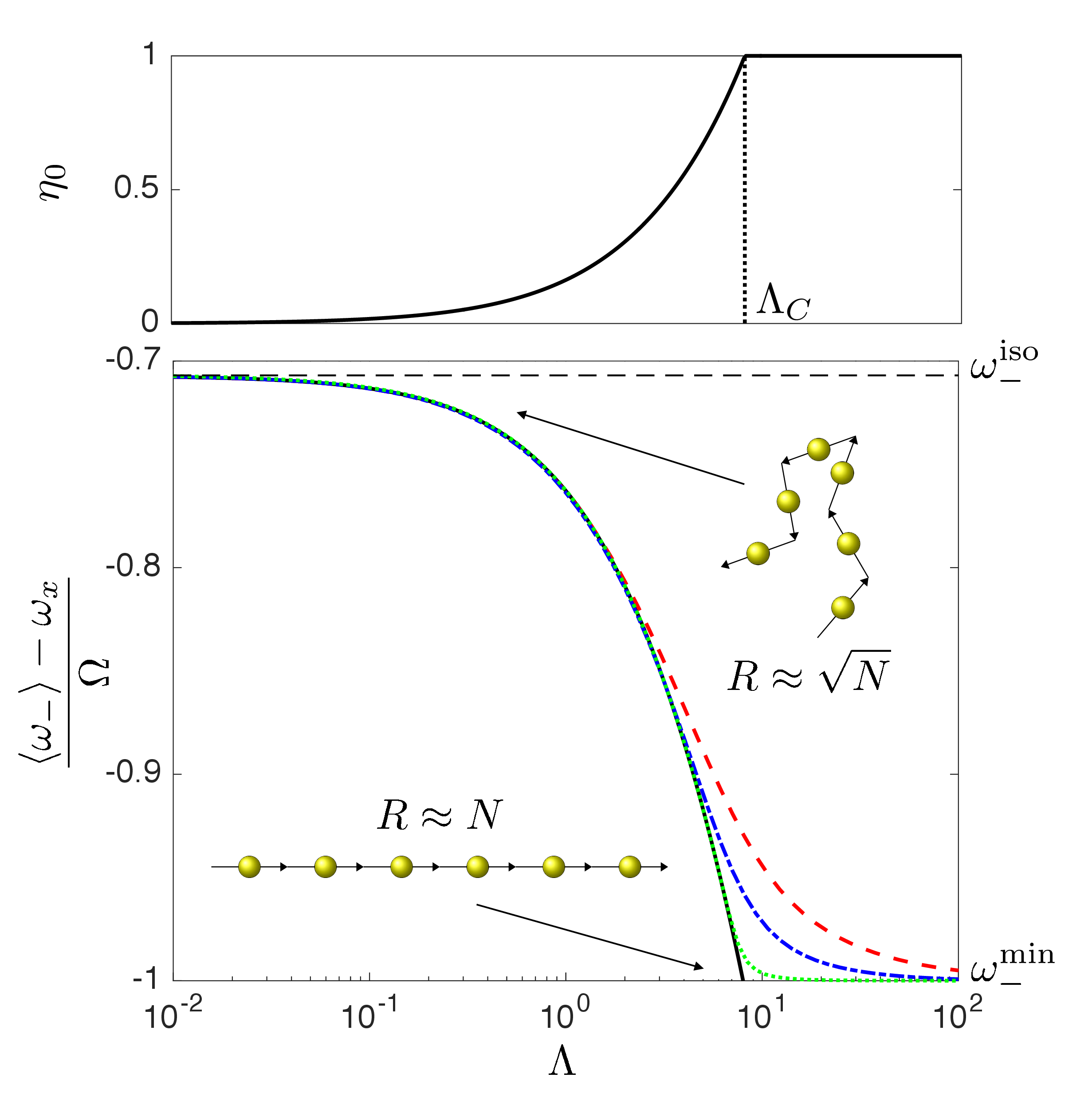}
	\caption{Top panel: $\eta_0$, as a function of the normalised inverse temperature $\Lambda$.
			Bottom panel: Energy of the lower polariton resonance as a function of $\Lambda$, obtained through the Laplace method in \Eq{wL} (black solid line), 
and numerically integrating \Eq{ZP} either with the exact form of the endpoint distribution from \Eq{Exact} for $N=10$ (red dashed line) and using the Gaussian form from \Eq{Gaussian} for $N=10$ (blue dash-dotted line) and $100$ (green dotted line).}\label{Fig2}
\end{figure}

In order to uncover the nature of the phase transition and to precisely identify the phase boundary we are led to calculate the partition function of the system with $M$ lower polaritons. Assuming the energy of such a state to be $M\omega_-$, an hypothesis we will critically assess in the last part of this Letter, the partition function reads
\begin{eqnarray}
\label{Z}
Z&\propto&\prod_{n=1}^N \int_0^{2\pi}d\theta_n \exp \left[M\beta \chi\sqrt{\sum_{n=1}^N \cos^2(\theta_n)}\right],
\end{eqnarray}
where $\beta=\frac{1}{k_BT}$ is the inverse temperature
and we choose $M\omega_x$ as energy reference.
In \Eq{Z} we neglected the upper polariton mode, since we are only interested in systems that verify $\beta \Omega > 1$, as otherwise the phase transition would be out of reach for any realistic value of $s$.
The nonlinear, collective interaction between the different dipoles in \Eq{Z} can be put in a more manageable form by introducing 
$N$ two-dimensional unit vectors $\mathbf{r}_n=\left[\cos(2\theta_n),\sin(2\theta_n)\right]$ and
noticing that 
\begin{eqnarray}
\sum_{n=1}^N \cos^2(\theta_n)&=&\frac{N}{2} + \frac{1}{2}\sum_{n=1}^N \cos(2\theta_n)=\frac{N+R_x}{2},
\end{eqnarray}
where $R_x$ is the $x$ component of the vector $\mathbf{R}=\sum_{n=1}^N\mathbf{r}_n$.
We can now recognise in \Eq{Z} the integral over the parameter space of a two-dimensional polymer made of $N$ segments of unit length, evolving in a potential depending on the total length of the chain along the $x$ direction. 
This identification of the set of dipoles with a polymer, sketched in Fig. \ref{Fig1}b, allows us to write the partition function as
\begin{eqnarray}
\label{ZP}
Z&\propto&\int_{\mathbb{R}^2} d^2\mathbf{R}\, P_N(R)\exp \left[\frac{s N \beta\chi}{\sqrt{2}}\sqrt{N+R_x}\right],
\end{eqnarray}
where $P_N(R)$ is the endpoint distribution of a two-dimensional polymer of length $N$ \cite{Kleinert} 
\begin{eqnarray}
\label{Exact}
P_N(R)= \int_0^\infty  dt \, t\, J(tR)\, J^N(R),
\end{eqnarray}
with $J$ the Bessel function of order zero. 
Numerical evaluation of \Eq{ZP} using the exact endpoint distribution becomes quickly infeasible for $N\gg1$, where instead we can rely on its Gaussian approximation
\begin{eqnarray}
\label{Gaussian}
P_N(R)\approx\frac{1}{\pi N}\exp\left(-\frac{R^2}{N}\right),
\end{eqnarray}
exact for $N\rightarrow\infty$ \cite{Kleinert}.
Using \Eq{Gaussian} into \Eq{ZP} we can evaluate the integral over $R_y$ and, introducing the normalised coordinate $\eta=R_x/N\in \left[0,1\right]$, obtain
\begin{eqnarray}
\label{ZG}
Z&\propto&\int_{0}^{1} d\eta \; \text{erf}\left[\sqrt{(1-\eta^2)N}\right] \,\exp\left[ Ng(\eta)\right],
\end{eqnarray}
where  $\text{erf}$ is the error function, $g(\eta)=\left[-\eta^2+ \Lambda\sqrt{\frac{1+\eta}{2}}\right]$, and $\Lambda=s\beta\Omega$ is the normalised inverse temperature. 
For $\Lambda<\Lambda_{C}=8$, $g(\eta)$ has an absolute maximum at 
$\eta_0\in(0,1)$, with $\Lambda_{C}$ corresponding to $\eta_0=1$, describing a fully stretched polymer. In the top panel of Fig. \ref{Fig2} $\eta_0$ is plotted as a function of the normalised inverse temperature $\Lambda$.
For $N\gg1$ and $\Lambda<\Lambda_{C}$ we can thus analytically calculate the partition function using the Laplace method, leading to
\begin{eqnarray}
\label{ZL}
Z\propto\sqrt{\frac{2\pi}{N g''(\eta_0)}}\quad \text{erf}\left[\sqrt{(1-\eta_0^2)N}\right] \exp \left[Ng(\eta_0)\right].
\end{eqnarray}
The average energy of the lower polariton resonance can now be calculated dividing the
expectation value of energy by the number of polaritons $M$ 
\begin{eqnarray}
\langle\omega_-\rangle=\omega_x-\frac{\Omega}{N}\frac{d\ln Z}{d \Lambda}, 
\end{eqnarray}
that from \Eq{ZL}, in the thermodynamic limit $N\rightarrow\infty$ with $\Omega$ finite, takes the form
\begin{eqnarray}
\label{wL}
\langle\omega_-\rangle^{\text{th}}=\omega_x- \Omega\sqrt{\frac{1+\eta_0}{2}}.
\end{eqnarray}
The average lower polariton energy thus interpolates between $\omega_-^{\text{iso}}$ and $\omega_-^{\text{min}}$ as the polymer passes from its bunched ($R\approx \sqrt{N}$, $\eta_0\approx 0$) to its stretched ($R\approx N$, $\eta_0\approx 1$) phase. 
\psfrag{OM}{$\frac{\langle\omega_-\rangle -\omega_x}{\Omega}$}
\psfrag{A}{a}	
In the bottom panel of Fig. \ref{Fig2} we plot the energy of the lower polariton resonance as a function of the normalised inverse temperature $\Lambda$, obtained through the Laplace method in \Eq{wL} (black solid line), 
and numerically integrating \Eq{ZP} with the exact form of the endpoint distribution from \Eq{Exact} for $N=10$ (red dashed line) and using the Gaussian from \Eq{Gaussian} for $N=10$ (blue dash-dotted line) and $100$ (green dotted line). We can see that the system undergoes a second order phase transition at the critical normalised temperature $\Lambda_{C}=8$, that is a critical temperature $k_BT_C=\frac{s\Omega}{8}$. 
Numerical and analytical results support our initial intuition. 
At high excitation density (low temperature, $\Lambda \geq 8$) the optomechanical cavity action is strong enough to overcome thermal fluctuations and align all the dipoles parallel to the electric field, thus increasing the coupling and pinning the polariton energy to its minimal value $\omega_-^{\text{min}}$ from \Eq{wmin}. 
At low excitation density instead (high temperature, $\Lambda\ll8$) the force exerted by the cavity field on each dipole is not enough to fight against thermal fluctuations and the dipoles end up in an isotropic configuration. When the dipoles are not aligned with the electric field their average coupling is smaller, and the polariton energy starts to increase, converging at the isotropic value $\omega_-^{\text{iso}}$ from \Eq{wiso} for 
$\Lambda\approx 0.1$.
The phase transition would thus manifest itself as an intensity-dependent shift of the lower polariton resonance upon optical pumping. 
In such a process only polaritonic bright states can be generated and thus the transition can be observed while the excitation density $s$ is kept constant through the interplay of pump and losses long enough for the system to thermalize, regardless of the polariton lifetime.

\begin{figure}
\centering
	\includegraphics[width=0.5\textwidth]{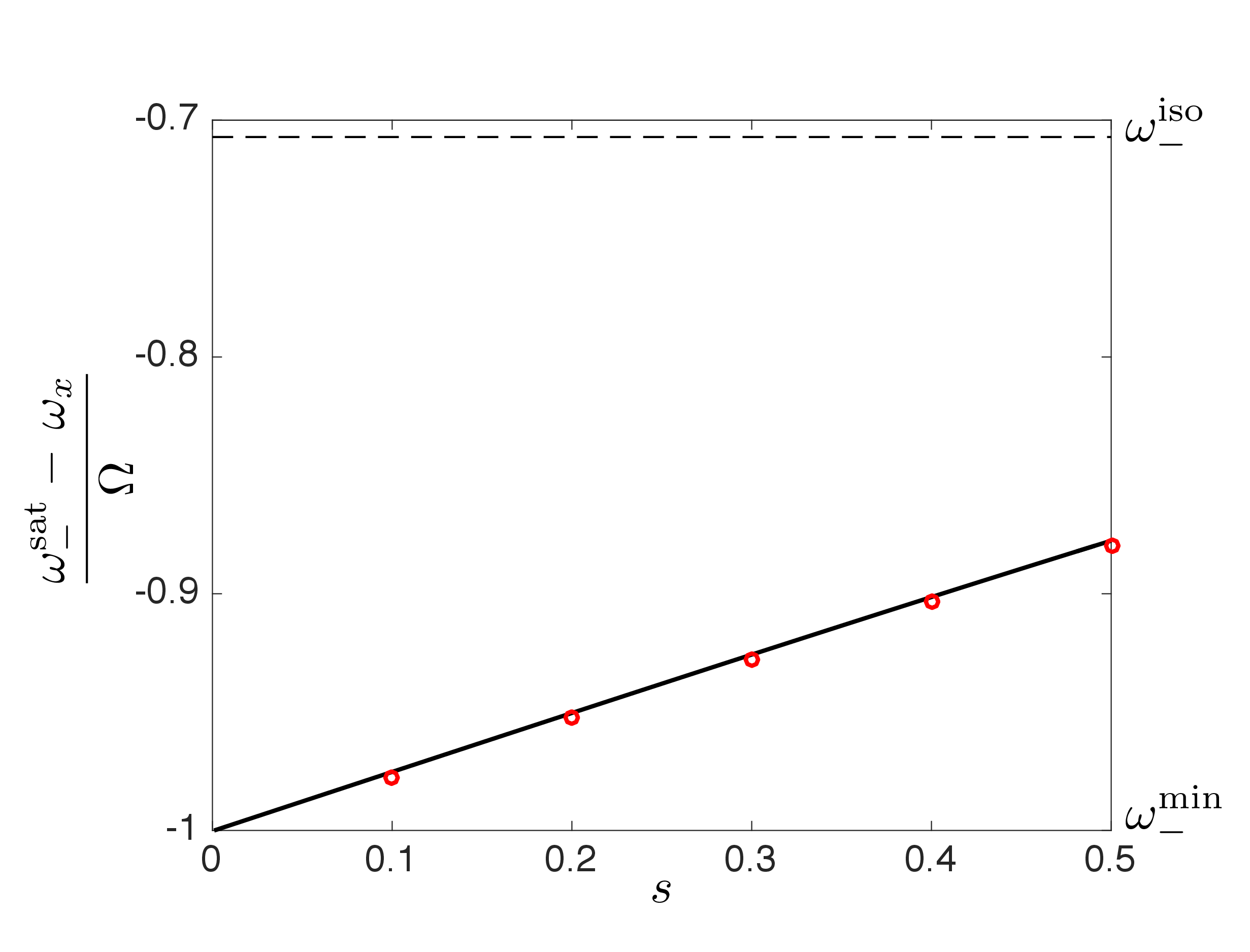}
	\caption{Plot as a function of the excitation density $s$ of the saturated lower polariton energy $\omega_-^{\text{sat}}$. 
	 The black solid line corresponds to $N=100$ while the red dots to $N=1000$. The black dashed line marks instead the isotropic value from the bosonic theory.}
	\label{Fig3}
\end{figure}

In writing \Eq{Z} we made the assumption that the energy of the lowest lying state in the $M$ excitations manifold can be written as $M$ times the lower polariton energy $\omega_-$ from \Eq{freq}. While this is certainly true for a bosonic (harmonic) system, polaritons are
strictly bosonic only in the dilute excitation regime $s\ll1$ and we thus need to assess up to which value of $s$ our theory remains accurate.
To do so we numerically diagonalise the Hamiltonian in \Eq{H} in the $M$ excitation manifold for the homogeneous case $\theta_n=0\;\forall \, n$. The lowest lying eigenvalue divided by $M$, that is the saturated lower polariton energy $\omega_-^{\text{sat}}$, is plotted in Fig. \ref{Fig3} as a function of the excitation density $s$. The results are plotted for $N=100$ (black solid line) and  $N=1000$ (red circles), showing that convergence has been achieved.The $\omega_-^{\text{sat}}$ shift with $s$ highlights physics beyond the present bosonic treatment, including possibly an increased impact of the RWA. Still, comparing $\omega_-^{\text{sat}}$ with the value $\omega_-^{\text{min}}$ expected for a perfectly bosonic system, we can see that saturation leads to changes sizeably smaller than the effect we expect to observe, and it can thus be neglected for not too large values of $s$.
\psfrag{sy}[bc][bc][1.6]{$\frac{\omega_-^{\text{sat}} -\, \omega_x}{\Omega}$}
\psfrag{sx}[bc][bc][1.2]{$s$}	
\psfrag{w1}[bl][bl][1.2]{$\omega_-^{\text{min}}$}	
\psfrag{w2}[bl][bl][1.2]{$\omega_-^{\text{iso}}$}	

Organic microcavities are a promising system for an experimental implementation of our model. In those systems
polariton condensates with filling fractions $s$ of the order of $0.1$ have been achieved \cite{Daskalakis14}, although we wish to stress that presence of coherence play no role in our theory. Note that the presence of a non-resonant pump used to create the condensate could also lead to spurious effects on the dipole orientation due to AC-Stark effect. Those nevertheless can be made negligibly small by increasing the detuning or completely removed using a non-polarised pump.

Record values of the light-matter coupling, up to $\Omega\approx 500$ meV, have been achieved in molecular crystals \cite{Schwartz11,Kena-Cohen13,Gambino14,Gubbin14}, that for an excitation density $s=0.2$ corresponds to a critical temperature $k_BT_C=150$ K, with the onset of the phase transition clearly visible well above room temperature.
Strong coupling has also been obtained in floating molecules and molecular liquids \cite{George15,George16}, although using vibrational and not electronic transitions, leading to much smaller couplings. 
Strongly coupled electronic transitions in floating organic molecules seem thus a natural candidate to observe the phase transition, where the presence of a viscous solvent would easily allow to satisfy the classicality and adiabaticity conditions we imposed on our model. 
The two-dimensional character of the dipoles could be implemented using phobic molecules floating on the surface of the solvent in an open cavity \cite{Schwarz14}, or a similar three-dimensional case could be realised with molecules in suspension using a flow-cell cavity, conceptually similar to the one described in Ref. \cite{George15}.

Our initial argument leads us to expect qualitatively the same physics
in the case of three-dimensional dipoles, with a transition between $\omega_-^{\text{min}}$ and a modified isotropic energy 
\begin{eqnarray}
\label{wiso3D}
\tilde{\omega}_{-}^{\text{iso}}&=&\omega_x-\chi\sqrt{\frac{N}{3}}=\omega_x-\frac{\Omega}{\sqrt{3}},
\end{eqnarray}
different from ${\omega}_{-}^{\text{iso}}$ because the average is now over the solid angle. Nevertheless it is not clear if the mapping between the set of dipoles and a polymer that allowed us to analytically solve the problem can be expanded to the three-dimensional case or if a different approach has to be developed, and we leave this for future investigations. 

In this Letter we proved that, exploiting highly excited states, collective interactions of dipoles with a photonic cavity can generate optomechanical effects allowing to manipulate the microscopic degrees of freedom associated with the individual dipoles. 
As a first example we studied the case of rotating, strongly coupled two-dimensional dipoles, proving a second order phase transition takes place, in which the system transitions from an isotropic to an aligned phase. This is only a first step in the investigation of the collective optomechanical effects in cavity quantum electrodynamics, and we expect the study of their thermodynamical properties will lead to a rich variety of new physics. Apart from the already discussed extension of the theory to the three-dimensional case, one venue that seems particularly promising for future investigations is the study of analogous processes realised with cold polar molecules in optical lattices. In those systems the classical and adiabatic approximations would not hold, and the dynamics of the internal degrees of freedom will thus play an important role, possibly leading to the appearance of novel, non-trivial phenomena.

\section{Acknowledgments}
The authors thank Johannes Feist for discussions and useful feedback.
S.D.L. is Royal Society Research Fellow and he acknowledges support from EPSRC grant EP/M003183/1.


\begin{thebibliography}{}

\bibitem{Dicke54} R. H. Dicke, \emph{Coherence in Spontaneous Radiation Processes}, Phys. Rev. {\bf 93}, 99 (1954). 

\bibitem{DeLiberato09} S. De Liberato and C. Ciuti, \emph{Stimulated Scattering and Lasing of Intersubband Cavity Polaritons}, Phys. Rev. Lett. {\bf 102}, 136403 (2009). 

\bibitem{Tuchman06} A. K. Tuchman, R. Long, G. Vrijsen, J. Boudet, J. Lee, and M. A. Kasevich, \emph{Normal-mode splitting with large collective cooperativity}, Phys. Rev. A {\bf 74}, 053821 (2006).
\bibitem{Arnold11} K. J. Arnold, M. P. Baden, and M. D. Barrett, \emph{Collective cavity quantum electrodynamics with multiple atomic levels}, Phys. Rev. A {\bf 84}, 033843 (2011).
\bibitem{Chen11} Z. Chen, J. G. Bohnet, S. R. Sankar, J. Dai, and J. K. Thompson, \emph{Conditional Spin Squeezing of a Large Ensemble via the Vacuum Rabi Splitting},
Phys. Rev. Lett. {\bf 106}, 133601 (2011).
\bibitem{Ningyuan16} J. Ningyuan, A. Georgakopoulos, A. Ryou, N. Schine, A. Sommer, and J. Simon, \emph{Observation and characterization of cavity Rydberg polaritons},
Phys. Rev. A {\bf 93}, 041802(R) (2016). 
\bibitem{Culver16} R. Culver, A. Lampis, B. Megyeri, K. Pahwa, L. Mudarikwa, M. Holynski, P. W. Courteille, and J. Goldwin, \emph{Collective strong coupling of cold potassium atoms in a ring cavity}, New J. Phys. {\bf 18}, 113043 (2016).




\bibitem{Weisbuch92} C. Weisbuch, M. Nishioka, A. Ishikawa, and Y. Arakawa, \emph{Observation of the coupled exciton-photon mode splitting in a semiconductor quantum microcavity},
Phys. Rev. Lett. {\bf 69}, 3314 (1992).
\bibitem{Lidzey98} D. G. Lidzey, D. D. C. Bradley, M. S. Skolnick, T. Virgili, S. Walker, and D. M. Whittaker, \emph{Strong exciton-photon coupling in an organic semiconductor microcavity}, Nature {\bf 395}, 53 (1998).
\bibitem{Dini03} D. Dini, R. K\"ohler, A. Tredicucci, G. Biasiol, and L. Sorba, \emph{Microcavity Polariton Splitting of Intersubband Transitions}, Phys. Rev. Lett. {\bf 90}, 116401 (2003).
\bibitem{Scalari12} G. Scalari, C. Maissen, D. Turcinkova, D. Hagenm\"uller, S. De Liberato, C. Ciuti, C. Reichl, D. Schuh, W. Wegscheider, M. Beck, and J. Faist, \emph{Ultrastrong Coupling of the Cyclotron Transition of a 2D Electron Gas to a THz Metamaterial}, Science {\bf 335}, 1323 (2012).
\bibitem{Zhang14} X. Zhang, C.-L. Zou, L. Jiang, and H. X. Tang, \emph{Strongly Coupled Magnons and Cavity Microwave Photons},
Phys. Rev. Lett. {\bf 113}, 156401 (2014).
\bibitem{Coles14} D. M. Coles, Y. Yang, Y. Wang, R. T. Grant, R. A. Taylor, S. K. Saikin, A. Aspuru-Guzik, D. G. Lidzey, J. Kuo-Hsiang Tang and J. M. Smith, \emph{Strong coupling between chlorosomes of photosynthetic bacteria and a confined optical cavity mode}, Nat. Comm. {\bf 5}, 5561 (2014).
 

\bibitem{Liu15} X. Liu, T. Galfsky, Z. Sun, F. Xia, E. Lin, Y-H. Lee, S. K\' ena-Cohen, and V. M. Menon, \emph{Strong light-matter coupling in two-dimensional atomic crystals}, Nat. Phot. {\bf 9}, 30 (2015). 
\bibitem{Gubbin16} A. Gubbin, F. Martini, A. Politi, S. A. Maier, and S. De Liberato, \emph{Strong and Coherent Coupling between Localized and Propagating Phonon Polaritons}, Phys. Rev. Lett. {\bf 116}, 246402 (2016).




\bibitem{Hutchison12} J. A. Hutchison, T. Schwartz, C. Genet, E. Devaux, and T. W. Ebbesen, \emph{Modifying chemical landscapes by coupling to vacuum fields}, Angew. Chem. Int. Ed. {\bf 51}, 1592 (2012).
\bibitem{Cwik14} J. A. \'Cwik, S. Reja, P. B. Littlewood, and J. Keeling, \emph{Polariton condensation with saturable molecules dressed by vibrational modes}, Eur. Phys. Lett. {\bf 105}, 47009 (2014).
\bibitem{Shalabney15}
A. Shalabney, J. George, J. Hutchison, G. Pupillo, C. Genet, and T. W. Ebbesen, \emph{Coherent coupling of molecular resonators with a microcavity mode}, Nat. Comm. {\bf 6}, 5981 (2015).


\bibitem{Galego16} J. Galego, F. J. Garcia-Vidal, and J. Feist,
\emph{Suppressing photochemical reactions with quantized light fields},
Nat. Comm. {\bf 7}, 13841 (2016).
\bibitem{Galego15} J. Galego, F. J. Garcia-Vidal, and J. Feist, \emph{Cavity-Induced Modifications of Molecular Structure in the Strong-Coupling Regime}, Phys. Rev. X {\bf 5}, 041022 (2015).
\bibitem{Cwik16} J. A. Cwik, P. Kirton, S. De Liberato, and J. Keeling, \emph{Excitonic spectral features in strongly-coupled organic polaritons}, Phys. Rev. A {\bf 93}, 033840 (2016).


\bibitem{Uphoff15} M. Uphoff, M. Brekenfeld, G. Rempe, and S. Ritter, \emph{Frequency splitting of polarization eigenmodes in microscopic FabryÐPerot cavities}, New J. Phys. {\bf 17}, 013053 (2015). 


\bibitem{Daskalakis14} K. S. Daskalakis, S. A. Maier, R. Murray, and S. K{\'e}na-Cohen, \emph{Nonlinear interactions in an organic polariton condensate}, Nat. Mater. {\bf 13}, 271 (2014).

\bibitem{Cummings83}
F. W. Cummings and A. Dorri, \emph{Exact solution for spontaneous emission in the presence of N atoms}, Phys. Rev. A {\bf 28}, 2282 (1983).

\bibitem{Lopez07}
C. E. L\'opez, F. Lastra, G. Romero, and J. C. Retamal, \emph{Entanglement properties in the inhomogeneous Tavis-Cummings model}, Phys. Rev. A {\bf 75}, 022107 (2007).

\bibitem{Emary13} C. Emary, \emph{Dark states in multi-mode multi-atom Jaynes-Cummings systems}, J. Phys. B: At. Mol. Opt. Phys. {\bf 46}, 224008 (2013).
\bibitem{Shammah16} N. Shammah and S. De Liberato, \emph{Superfluorescence in presence of strong dephasing in solid state systems}, To be submitted.

\bibitem{Kleinert} H. Kleinert, \emph{Path Integrals in Quantum Mechanics, Statistics, Polymer Physics, and Financial Markets}, World Scientific, Singapore 2009.

\bibitem{Gambino14}
S. Gambino, M. Mazzeo, A. Genco, O. Di Stefano, S. Savasta, S. Patan\' e, D. Ballarini, F. Mangione, G. Lerario, D. Sanvitto, and G. Gigli, \emph{Exploring Light-Matter Interaction Phenomena under Ultrastrong Coupling Regime}, ACS Photonics {\bf 1}, 1042 (2014).

\bibitem{Schwartz11}
T. Schwartz, J. A. Hutchison, C. Genet, and T. W. Ebbesen, \emph{ Reversible Switching of Ultrastrong Light-Molecule Coupling}, Phys. Rev. Lett. {\bf 106}, 196405 (2011).

\bibitem{Kena-Cohen13} 
S. K\' ena-Cohen, S. A. Mayer, and D. D. C. Bradley,\emph{Ultrastrongly Coupled Exciton-Polaritons in Metal-Clad Organic Semiconductor Microcavities}, Advanced Optical Materials {\bf 1}, 827 (2013).

\bibitem{Gubbin14}
C. R. Gubbin, S. A. Maier, and S. K\' ena-Cohen, \emph{Low-voltage polariton electroluminescence from an ultrastrongly coupled organic light-emitting diode}, Appl. Phys. Lett. {\bf 104}, 233302 (2014).


\bibitem{George15} J. George, A. Shalabney, J. A. Hutchinson, C. Genet, and T. W. Ebbesen, \emph{Liquid-Phase Vibrational Strong Coupling}, J. Phys. Chem. Lett. {\bf 6}, 1027 (2015).
\bibitem{George16} J. George, T. Chervy, A. Shalabney, E. Devaux, H. Hiura, C. Genet, and T. W. Ebbesen, \emph{Multiple Rabi Splittings under Ultrastrong Vibrational Coupling}, Phys. Rev. Lett. {\bf 117}, 153601 (2016).


\bibitem{Schwarz14} S. Schwarz, S. Dufferwiel, P. M. Walker, F. Withers, A. A. P. Trichet, M. Sich, F. Li, E. A. Chekhovich, D. N. Borisenko, N. N. Kolesnikov, K. S. Novoselov, M. S. Skolnick, J. M. Smith, D. N. Krizhanovskii, and A. I. Tartakovskii, \emph{Two-Dimensional MetalÐChalcogenide Films in Tunable Optical Microcavities}, Nano Lett. {\bf 14}, 7003 (2014).


\end{thebibliography}
\end{document}